\documentclass[twocolumn,english,pre,superscriptaddress]{revtex4}
\usepackage[latin9]{inputenc}
\usepackage{amsbsy}
\usepackage{graphicx}
\usepackage{amssymb}

\makeatletter
\@ifundefined{definecolor}
 {\usepackage{color}}{}

\makeatother

\usepackage{babel}

\begin{document}

\title{ A thermodynamic force generated by chemical gradient and adsorption reaction }

\author{Takeshi Sugawara}

\affiliation{Department of Basic Science, Graduate School of Arts and Science,
\\
 The University of Tokyo, 3-8-1, Komaba, Meguro, Tokyo 153-8902,
Japan}

\author{Kunihiko Kaneko}

\affiliation{Department of Basic Science, Graduate School of Arts and Science,
\\
 The University of Tokyo, 3-8-1, Komaba, Meguro, Tokyo 153-8902,
Japan}

\affiliation{Complex Systems Biology Project, ERATO, JST, Komaba, Meguro-ku, Tokyo,
Japan}

\begin{abstract}
Biological units such as macromolecules, organelles, and cells are directed to a proper location under gradients of relevant chemicals.
By considering a macroscopic element that has binding sites for a chemical adsorption reaction to occur on its surface, we show the existence of a thermodynamic force that is generated by the gradient and exerted on the element. 
 By assuming local equilibrium and adopting the grand potential from thermodynamics, we derive a formula for such a thermodynamic force, which depends on the chemical potential gradient and Langmuir isotherm. 
The conditions under which the formula can be applied are demonstrated to hold in intracellular reactions. 
The role of the force in the partitioning of bacterial chromosome/plasmid during cell division is discussed.
\end{abstract}
\maketitle

\paragraph*{Introduction  }

Biological entities such as cells, organelles, and macromolecules
often move to a proper location under an external gradient of chemicals.
In the chemotaxis of a cell, the process by which an organism senses
the presence of an external chemical and responds to it has been elucidated
in depth. However, such coordinated motion is not limited to the organism
level. In recent studies, it has been revealed that such directional
motion under chemical gradients/localizations plays an important
role in organization at an intracellular level \cite{gradient}, e.g.,
microtubule guidance under the RanGTP gradient \cite{microtubule-1} or Stathmin gradient \cite{microtubule-3}, actin nucleation under the IcsA gradient on the outer membrane of
the pathogen \em Shigella flexneri \rm (actin comet)\cite{IcsA-gradient}, bacterial chromosome/plasmid DNA partitioning by a Par system \cite{caulobacter1,caulobacter2,chorelae1,Par-system1,F-plasmid-positioning-0,F-plasmid-positioning-1,F-plasmid-positioning-1.5,F-plasmid-positioning-2}, and several transport processes by scaffold proteins. However, the general mechanism by
which organelles or macromolecules demonstrate coordinated motion
under an intracellular gradient/localization is yet to be determined.

The pattern formation process of chemical concentrations under a non-equilibrium
condition, first reported for a macroscopic morphogenesis\cite{Turing},
has recently been observed at an intracellular level as well, and
its relevance to intracellular organization processes has been demonstrated
in  
 bacterial plasmid DNA partitioning Par system\cite{Par-system1}\cite{F-plasmid-positioning-1},
 Min system for determining the plane of cell division in bacteria \cite{bacterial-cell-division-theore-0}, etc. They discussed how positional information given by the chemical
concentration gradient/localization is obtained and stored with the
mechanism. However, there are few studies on how the coordinated
motion or transport of organelles or macromolecules caused by chemical
gradient/localization is sustained under non-equilibrium conditions.

In the present paper, we present a physical mechanism of such directed
motion by showing that under a chemical gradient corresponding to
a reaction, a macroscopic element consisting of a number of reaction
sites is generally subject to a force. By representing this element
as a scaffold for the adsorption of a chemical, we derive a general
formula for a force generated by a chemical potential gradient. In
the derivation, we introduce the grand potential in thermodynamics
for an open system \cite{callen-text} and apply the second law of thermodynamics. Irrespective
of the nature of the molecules, the force acts to decrease the grand
potential when a chemical potential gradient exists. Then, the element
moves in a direction such that the chemical potential increases. A
formula is obtained by assuming that the reaction process reaches
equilibrium faster than the motion of the element and by extending
the minimization of free energy to include the contact with a particle
bath with a given chemical gradient. We propose that this force generates
a general mechano-chemical coupling driven by the chemical potential
gradient through an adsorption reaction.
The proposed formula is valid since the work
extracted by the gradient force dominates the thermal energy fluctuation.
 By examining whether this condition is satisfied in the intracellular
reaction process, we discuss the possible role of this force in the
partitioning dynamics of bacterial chromosome/plasmid for cell division.

\paragraph*{Chemical gradient force} 

\begin{figure}[htbp]
\begin{centering}
\includegraphics[scale=0.46]{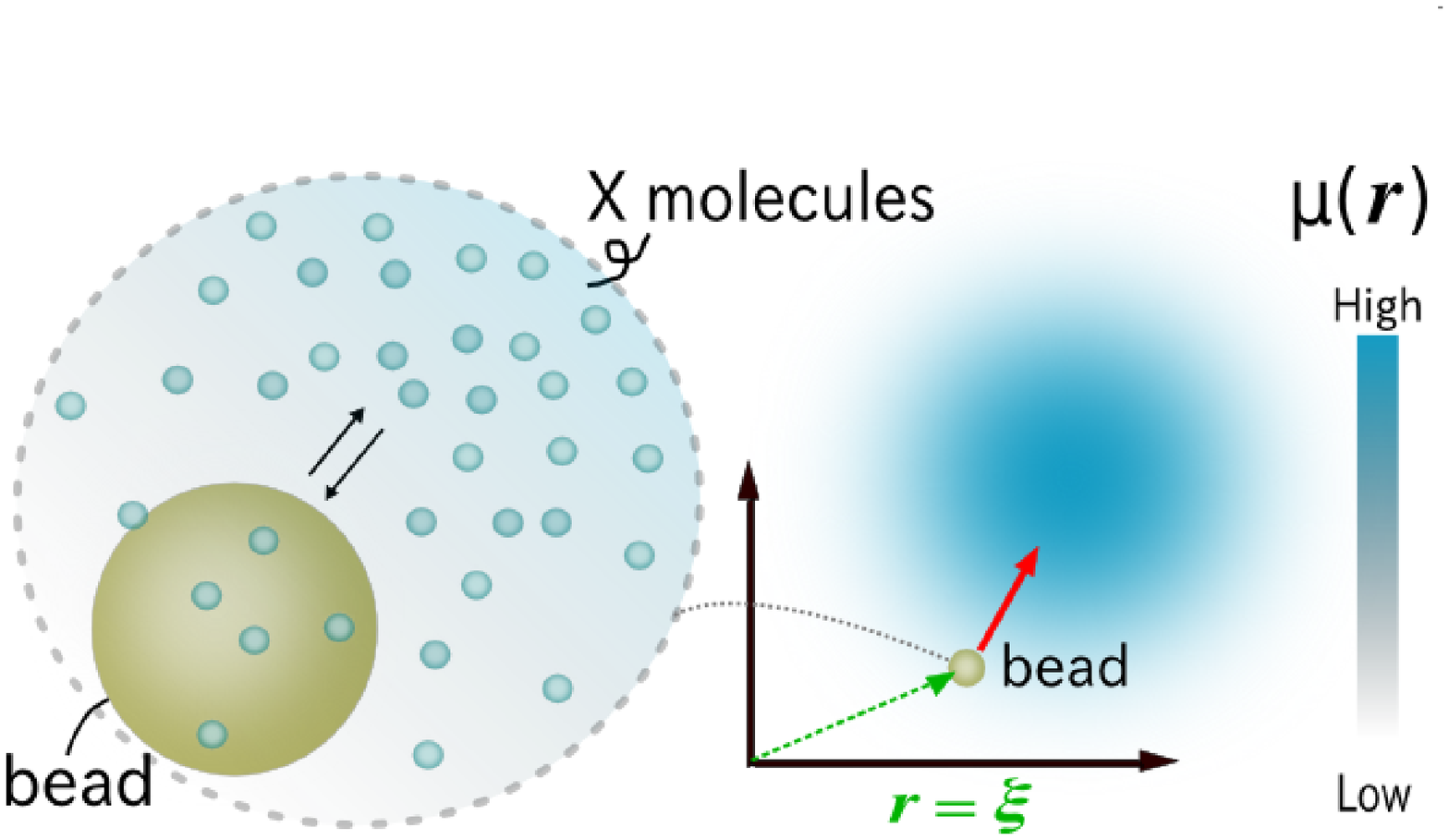}
\par\end{centering}

\caption{ Schematic representation of our system. A {}``bead'' is placed
at $\boldsymbol{r}=\boldsymbol{\xi}$ and moves in a $d$-dimensional
space ($d$=1,2,3). The adsorption reaction $\mathrm{X}(\boldsymbol{\xi})+\mathrm{B}\leftrightarrows\mathrm{Y}$
occurs on the surface of the bead.}

\centering{}\label{Fig1}
\end{figure}

\label{sec2} Now, consider organelles or macromolecules that have
a number of binding sites for reactions to occur on their surface;
for example, nucleoprotein complexes (NCs) have several promoter sites
to bind transcription factors, etc. Let us model these biological
elements simply as {}``beads'' with several reaction sites to which
the corresponding molecules attach, as shown in Fig.\ref{Fig1}. The
bead is placed at $\boldsymbol{r}=\boldsymbol{\xi}$ and moves in
a $d$-dimensional space $\boldsymbol{r}\in\boldsymbol{R}^{d}$ ($d$=1,2,3).
We consider an isothermal process that is homogeneous over space at
a given temperature $T$ , whereas we consider a chemical bath with
a spatially dependent concentration $x(\boldsymbol{r})$ of a chemical
X, or equivalently, the corresponding chemical potential $\mu(\boldsymbol{r})$.
This gradient is assumed to be sustained externally. This chemical
is attached to the binding sites B on the bead and thus forms a complex
Y (Fig.\ref{Fig1}), as given by the reaction $\hspace{.05in}{\rm X(\boldsymbol{\xi})+B\leftrightarrows Y}$.
The molecular number of the complexes on the bead is denoted by $N_{y}$.
Note that the {}``bead'' here is defined as a macroscopic entity
compared to the molecule X. To develop local-equilibrium conditions,
we make the following assumptions. The {}``adsorption'' reaction
on the bead is considered to be a macroscopic event and $N_{y}$ is
a molecular number averaged over a much longer timescale
than the microscopic timescale of the reaction. Further, the bead
is assumed to move sufficiently slowly so that local chemical equilibrium
of the above reaction can be assumed to exist at each position $\boldsymbol{r}=\boldsymbol{\xi}$.
In other words, the timescales of diffusion of X molecules $\tau_{diff}$
and adsorption reaction $\tau_{adsorb}$ are much smaller than the
timescale of the motion of the bead $\tau_{bead}$: $\tau_{diff},\tau_{adsorb}\ll\tau_{bead}$.
As long as the bead is in motion, it is maintained in equilibrium
with the reservoir at $\boldsymbol{r}=\boldsymbol{\xi}$.

With the assumption of the existence of local equilibrium, we can
apply thermodynamics with spatially dependent thermodynamic variables.
Indeed, at each position $\boldsymbol{\xi}$, the reaction process
is described by the familiar classical Langmuir adsorption theory \cite{physcell}\cite{langmuir};
Note that the theory has been successfuly applied to DNA-protein binding equilibrium \cite{physcell}. 
The grand potential in thermodynamics at each position is given by $\hspace{.1in}\Omega(\boldsymbol{\xi})=F(\boldsymbol{\xi})-y\mu(\boldsymbol{\xi}),\hspace{.1in}d\Omega(\boldsymbol{\xi})=dF(\boldsymbol{\xi})-d(y\mu(\boldsymbol{\xi}))=-yd\mu(\boldsymbol{\xi})$,
where $y=\frac{N_{y}}{V}$ ($V:$ volume of the bead), $F(\boldsymbol{\xi})$
is the Hermholtz free energy, and $dF(\boldsymbol{\xi})=\mu(\boldsymbol{\xi})dy$ \cite{callen-text}\footnote{Here we assume that the gradient of X is not influenced by the adsorption reaction.
 This assumption is valid if $x(\boldsymbol{\xi})\gg y $.}.

Now, consider a virtual displacement for the position of the bead
$\boldsymbol{r}=\boldsymbol{\xi}$. Under an infinitesimal displacement
$\boldsymbol{\xi}\rightarrow\boldsymbol{\xi}+d\boldsymbol{\xi}$,
the change in the grand potential is $\hspace{.05in}d\Omega(\boldsymbol{\xi})=-yd\mu(\boldsymbol{\xi})=-y\boldsymbol{\nabla}\mu(\boldsymbol{\xi})\cdot d\boldsymbol{\xi}$.
In other words, the position of the bead $\boldsymbol{\xi}$ is adopted
as an effective independent variable instead of the chemical potential
$\mu(\boldsymbol{\xi})$. 
 Then, $\boldsymbol{\xi}$ is the work coordinate, while $-y\boldsymbol{\nabla}\mu(\boldsymbol{\xi})$ is the force exerted on the bead by the external world balanced by the force exerted on the bead due to reservoir chemical potential distribution $\mu(\boldsymbol{r})$.
 The chemical gradient force generated by the reservoir acts on the bead per unit
volume of the bead: \begin{equation}
\boldsymbol{f}_{chem}=-\boldsymbol{\nabla}\Omega(\boldsymbol{\xi})=y\boldsymbol{\nabla}\mu(\boldsymbol{\xi})\end{equation}

This expression is interpreted as follows: consider a quasi-static
infinitesimal displacement $d\boldsymbol{\xi}$. Then, from the change
in the grand potential, the maximum work that the system does on the
external world through the reservoir is $d^{\prime}W=y\boldsymbol{\nabla}\mu(\boldsymbol{\xi})\cdot d\boldsymbol{\xi}$.
By considering that this work is due to the force that the reservoir
exerts on the bead, as $d^{\prime}W=\boldsymbol{f}_{chem}\cdot d\boldsymbol{\xi}$,
the force formula eq.(1) is obtained. In other words, without an externally
applied force, $\boldsymbol{\xi}$ evolves spontaneously so that $\Omega(\boldsymbol{\xi})$
monotonically decreases, that is, $d\Omega(\boldsymbol{\xi})<0$ 
\footnote{The variational principle for the grand potential $\Omega$ can be
proved by the principle of maximum work in an isothermal process.
See Supplementary Information for a detailed derivation.}. 
When we consider the overdamped system, where the kinetic energy
of the bead is negligible, the equation of motion is given by 
$-\dot{\Omega}(\boldsymbol{\xi})=T\sigma(\boldsymbol{\xi})\hspace{.02in}>0$,
where $\sigma(\boldsymbol{\xi}):=\frac{d_{ir}S}{dt}(\boldsymbol{\xi})$
is the entropy production of the system. The above equation of motion
is rewritten as $\gamma\hspace{.05in}\boldsymbol{\dot{\xi}}=-\boldsymbol{\nabla}\Omega(\boldsymbol{\xi})=y\boldsymbol{\nabla}\mu(\boldsymbol{\xi})$,
when there is dissipation due to friction, assumed to be proportional 
to the velocity with the proportionality constant $\gamma$---the
friction coefficient per unit volume of the bead. 

Now, consider the condition for local chemical equilibrium. Because
$c=y+b=const.$, the dissociative constant is defined as $K:=\frac{k_{-}}{k_{+}}=\frac{x(\boldsymbol{\xi})b}{y}$,
where $y$ is given by the Langmuir isotherm $\hspace{.05in}y=y(x(\boldsymbol{\xi}))=c\hspace{.02in}\frac{x(\boldsymbol{\xi})}{K+x(\boldsymbol{\xi})}\hspace{.05in}$
\footnote{In terms of statistical mechanics, the Langmuir isotherm should be
obtained as a function of the chemical potential $\mu(\boldsymbol{r})$
at $\boldsymbol{r}=\boldsymbol{\xi}$. We can derive the representation
as a function of $\mu(\boldsymbol{r})$ when $\mu(\boldsymbol{r})$ is that for a dilute solution, i.e., $\mu(\boldsymbol{r})=\bar{\mu}+k_{B}T\ln x(\boldsymbol{r})$.
Here, $\bar{\mu}$ is the standard chemical potential.
}. If we consider cooperative adsorption of $n$ chemicals on the bead:
${\rm nX(\boldsymbol{\xi})+B\leftrightarrows Y}$, the isotherm for
the adsorption is given as $y(x(\boldsymbol{\xi}))=c\hspace{.02in}\frac{x(\boldsymbol{\xi})^{n}}{K^{n}+x(\boldsymbol{\xi})^{n}}$,
where $n$ determines a Hill coefficient. Therefore, the equation
of motion is written as \begin{equation}
\gamma\hspace{.05in}\boldsymbol{\dot{\xi}}=c\hspace{.02in}\frac{x(\boldsymbol{\xi})^{n}}{K^{n}+x(\boldsymbol{\xi})^{n}}\boldsymbol{\nabla}\mu(\boldsymbol{\xi})\end{equation}

Equation (2) is a general formula for the motion of an element that has a number of binding sites for chemical adsorption under the gradient of corresponding chemical potential.
 The theoretical description of the motion of the bead is somewhat an extended form of Langmuir adsorption theory. 
Because of this force, the bead moves in a direction such that the grand potential $\Omega$, the thermodynamic potential for open systems, is decreased; this decrease follows from the second law of thermodynamics.
Accordingly, the bead moves in the direction of increasing chemical potential.

In our formulation, the chemical potential $\mu(\boldsymbol{r})$
is regarded to be equivalent to the external potential field for bead
motion, similar to the magnetic field in which a magnetic dipole is placed. 
Indeed, the magnetic force that
acts on the dipole placed in the magnetic field gradient acts in the
direction of increasing magnetic field strength.

When an additional external potential field $E(\boldsymbol{r})$ such as elastic energy
 is applied to the bead, the relation
$dF(\boldsymbol{\xi})=dE(\boldsymbol{\xi})+\mu(\boldsymbol{\xi})dy(\boldsymbol{\xi})$
holds from the first law of thermodynamics. Accordingly, the grand
potential of the system given by $\Omega(\boldsymbol{\xi})=F(\boldsymbol{\xi})-y(\boldsymbol{\xi})\mu(\boldsymbol{\xi})$
satisfies $d\Omega(\boldsymbol{\xi})=dF(\boldsymbol{\xi})-d\left(\mu(\boldsymbol{\xi})y(\boldsymbol{\xi})\right)=dE(\boldsymbol{\xi})-y(\boldsymbol{\xi})d\mu(\boldsymbol{\xi})$.
Further, by taking into account thermal fluctuation, the equation
of motion of the system is given by 
\begin{equation}
\gamma\hspace{.05in}\boldsymbol{\dot{\xi}}=-\boldsymbol{\nabla}E(\boldsymbol{\xi})+y(\boldsymbol{\xi})\boldsymbol{\nabla}\mu(\boldsymbol{\xi})+\boldsymbol{\eta}(t)
\end{equation}
 with $\left<\boldsymbol{\eta}(t)\right>=0,\hspace{.05in}\left<\boldsymbol{\eta}(t)\boldsymbol{\cdot}\boldsymbol{\eta}(t^{\prime})\right>=2d\gamma k_{B}T\delta(t-t^{\prime})$.
Note that the equation is obtained as an extended form of a conventional
one with the Hermholtz free energy as the thermodynamic potential
for a closed system. It is straightforward to further extend the present
formula to include a case involving multiple beads and multiple components
with interactions among the beads.

\paragraph*{An example  }

We consider a simple toy model as an example of eq.(3).
We assume that a bead placed at $r=\xi$ is balanced at $r=0$ in a 
1-dimensional space $r$ by a restoration force, which is produced
by a linear spring and represented by a harmonic potential $E(\xi)=\frac{a}{2}\xi^{2}$.
The concentration of the chemical that reacts with the bead is assumed
to be sustained and is given by $x(r)=x_{s}\exp(\lambda(r-r_{s}))$,
where $x_{s}=$$x(r_{s})$ is the concentration of X at $r=r_{s}$,
 while the corresponding chemical potential is given by $\mu(r)=\bar{\mu}+k_{B}T\ln x(r)$.
Here, $\bar{\mu}$ is the standard chemical potential. Then, by using
eq.(3) without considering thermal fluctuation, the equation of motion
of the bead is obtained as $\hspace{.05in}\gamma\hspace{.05in}\dot{\xi}=y(\xi)\frac{d\mu}{dr}(\xi)-\frac{dE}{dr}(\xi)=\lambda k_{B}Tc\frac{x(\xi)^{n}}{K^{n}+x(\xi)^{n}}-a\xi$.
Here, $\gamma,c,K,a$ are the frictional coefficient, maximum adsorption concentration,
dissociative constant, and spring constant, respectively. This equation
can be rewritten as $\hspace{.05in}\gamma\hspace{.05in}\dot{\xi}=\frac{A}{1+B\exp(-n\lambda(\xi-r_{s}))}-a\xi$
with $A=\lambda k_{B}Tc,B=\left(\frac{K}{x_{s}}\right)^{n}$. The
solution of the equation shows bistability as it includes a sigmoid
function. Thus, under appropriate parameter regions, as $x_{s}$,
which is taken as a control parameter, is increased, the model thus
undergoes two-fold bifurcations from monostable fixed points to bistable
ones and then back to monostable states. This implies that the gradient
force determines the stable fixed position of the bead. 
 This toy model can be used to explain how the stable position of
an organella undergoes bifurcation to a new location as the chemical
concentration increases. An example may be seen in bacterial chromosome/plasmid
partitioning, as will be discussed later.

\paragraph*{Conditions for  chemical gradient force (i) : relationships among timescales }

The grand canonical description with local chemical equilibrium can
be adopted only in certain conditions. 
The following relationships among timescales must exist: (1) the timescale of adsorption
$\tau_{adsorb}$ must be shorter than that of bead motion $\tau_{bead}$
in order for the adsorption reaction to reach equilibrium (approximately)
before the bead moves, and (2) the timescale of bead motion must be
much larger than that of the diffusion of the chemical adsorbed on
the bead, i.e., $\tau_{bead}\gg\tau_{diff}$. This condition is necessary
to assume macroscopic motion of the bead under the chemical gradient,
which itself diffuses in space. In the living cells of particular
interest, the molecular number is not so large that the instantaneous
chemical concentration fluctuates greatly. Hence, in order to determine
the macroscopic concentration, the temporal average of the concentration
of molecules over the diffusion timescale must be calculated \cite{physical-limit-due-to-noise-7},
and hence, the above condition is important.

\paragraph*{Condition (ii): dominance of chemical gradient force over thermal fluctuation }

As long as the above conditions are satisfied, on an average, the
gradient force acts in the direction in which the chemical potential
increases, regardless of thermal fluctuation. However, the gradient
force must be larger than thermal noise in order for the force to
act effectively without the need for long-time averaging to remove
fluctuation. This establishes the following condition: the work done
by the force must be larger than the thermal energy, i.e., $-V\int_{\boldsymbol{\xi}_{0}}^{\boldsymbol{\xi}}d\boldsymbol{\xi}^{\prime}y(\boldsymbol{\xi}^{\prime})\boldsymbol{\nabla}\mu(\boldsymbol{\xi}^{\prime})\hspace{.05in}\gtrsim\hspace{.05in}k_{B}T$
for directional motion from $\boldsymbol{\xi}_{0}$ to $\boldsymbol{\xi}$
\footnote{This inequality is also derived from the steady solution of the Smoluchowski equation. For one-dimensional motion, the steady-state distribution
of $\xi$ is given by $P(\xi)=A\exp\left(-\frac{W(\xi)}{k_{B}T}\right)$.
Here, $A$ is a normalized factor and $\hspace{.05in}W(\xi)=-V\int_{0}^{\xi}d\xi^{\prime}y(\xi^{\prime})\nabla\mu(\xi^{\prime})$.
If the position $\xi_{c}$ that satisfies $W(\xi_{c})\sim k_{B}T$
 is smaller than the system size $L$, the distribution is concentrated
toward higher chemical potentials and the gradient force dominates
the thermal fluctuation. Then, the inequality in the text is obtained.
}. In the case of one-dimensional motion from $r=0$ to $L$, the above
condition for the work is rewritten as follows: the concentration
and chemical potential of chemical X are obtained as $x(r)=x_{0}\exp(-\frac{r}{r_{c}})$
and $\mu(r)=\bar{\mu}+k_{B}T\ln x(r)$, respectively. Here, $x_{0}$
gives the highest concentration at $r=0$. Then, the above inequality
is written as $\frac{N_{c}}{r_{c}}\int_{0}^{L}d\xi\frac{x_{0}^{n}e^{-n\frac{\xi}{r_{c}}}}{K^{n}+x_{0}^{n}e^{-n\frac{\xi}{r_{c}}}}\gtrsim1$
with $N_{c}=Vc$. A straightforward calculation leads to the inequality
$\hspace{.05in}\frac{N_{c}}{n}\ln\left(\frac{1+\left(\frac{x_{0}}{K}\right)^{n}}{1+\left(\frac{x_{0}}{K}\right)^{n}\exp{\left(-n\frac{L}{r_{c}}\right)}}\right)\gtrsim1$.
We can rewrite the above inequality in the form $\frac{x_{0}}{K}\gtrsim\left(\frac{\exp(\frac{n}{N_{c}})-1}{1-\exp\left(-n\left(\frac{L}{r_{c}}+\frac{1}{N_{c}}\right)\right)}\right)^{\frac{1}{n}},$
with the additional condition $N_{c}>\frac{r_{c}}{L}$ 
\footnote{We can also represent the condition in terms of the average concentration
$x_{av}$ for $[0,L]$.
}. Note that the the lower bound on $\frac{x_{0}}{K}$ decreases with
$\frac{L}{r_{c}}$ and $N_{c}$, and the formula is valid over a wide range of $\frac{x_{0}}{K}$.

\paragraph*{An application }

Now, we consider applications of the chemical gradient force within a cell. As an example, consider the partitioning
of chromosome/plasmid in bacteria during cell division \cite{caulobacter1,caulobacter2,chorelae1,Par-system1,F-plasmid-positioning-0,F-plasmid-positioning-1,F-plasmid-positioning-1.5,F-plasmid-positioning-2}.
Here, the bead corresponds to the chromosome/plasmid on which the
relevant protein (X in the model) binds to form a nucleoprotein complex
(NC) (Y in the model) that is important for partitioning. For the
application, the above-mentioned condition on the timescales, $\tau_{diff},\tau_{adsorb}\ll\tau_{bead}$
has to be satisfied. 
In general, the timescale
of protein binding equilibrium on bacterial DNA $\tau_{adsorb}$ is
at most $\tau_{adsorb}\lesssim1(s)$. In \cite{macrodomainmobility},  
it has been suggested that the segregation of chromosomes is spatially
restricted and the diffusion coefficient $D$ is estimated as $O\left(10^{-4}\sim10^{-5}\right)$. 
The size of NC, $a$, is assumed to be on the order of $a\sim50(nm)$,
so that $\tau_{bead}$ is estimated as $\tau_{bead}\sim\frac{a^{2}}{D}\sim\frac{{0.05}^{2}}{10^{-4}}\sim25(s)$.
Similarly, $\tau_{diff}$ is estimated as $\tau_{diff}\sim\frac{a^{2}}{D_{x}}\sim10^{-3}(s)$,
where $D_{x}$ is the diffusion coefficient of proteins within cytoplasm and is roughly estimated as $D_{x}\sim3.0(\mu m^{2}/s)$
 (according to \cite{S.Xie}). Therefore, $\tau_{diff}\ll\tau_{adsorb}\ll\tau_{bead}$
is satisfied for our formula.

To be specific, the replicated bacterial chromosomes/plasmids are
actively partitioned to daughter cells by a Par system \cite{caulobacter1,caulobacter2,chorelae1,Par-system1,F-plasmid-positioning-0,F-plasmid-positioning-1,F-plasmid-positioning-1.5,F-plasmid-positioning-2}.
The most ubiquitous partitioning system is the $parABS$ system.
 Before cell division, daughter chromosomes/plasmids are precisely
segregated by ParA dynamic localization along with the associated
ParA-NC interaction. It has been suggested that ParA localization
pulls chromosomes\cite{chorelae1} or at least guides plasmid\cite{F-plasmid-positioning-2},
but how ParA generates the driving force for chromosome/plasmid segregation
has not been elucidated yet. Now, ParA is regarded as X, NC (to be
precise, ParB-$parS$  NC mediated by ParA ) as Y in our model 
\footnote{The Par system consists of three components: DNA binding protein
ParB, ATPase ParA, and centromere-like site $parS$. ParB binds
 $parS$, spreads along DNA, and forms a large NC around $parS$. ParA
can nonspecifically bind to DNA and interact with the ParB-$parS$ 
NC \cite{caulobacter1,caulobacter2,chorelae1,Par-system1,F-plasmid-positioning-0,F-plasmid-positioning-1,F-plasmid-positioning-1.5,F-plasmid-positioning-2}.
}; this driving force is naturally explained by the chemical gradient force. 
To confirm the validity of our formula, we simply consider the steady
concentration distribution of ParA to be $x(r)=x_{0}\exp(-\frac{r}{r_{c}})$
within $[0,L]$ \cite{caulobacter2,chorelae1}.  
We then take $n=2$, because ParA cooperatively binds DNA \cite{BacillusEMBO05} and choose
$K=0.3(\mu M)$ on the basis of recent data obtained by \cite{BacillusEMBO05}.
Although the precise number of ParA that binds DNA around NC is unknown,
$N_{c}\sim10$ is a natural choice since many ParA are suggested to
bind DNA around NC. The localization scale $r_{c}$ of ParA localization
is also not precisely known and could vary across bacterial species.
 By taking $\frac{L}{r_{c}}\sim5$ as an order estimate from \cite{caulobacter2,chorelae1},
we obtain $\frac{x_{0}}{K}\gtrsim0.5$.   
 Active protein binding generally occurs when a reasonable concentration
$x_{0}\sim K$ is reached, where $\frac{x_{0}}{K}$ is sufficiently
large to satisfy the above inequality. Therefore, the chemical gradient force 
dominates the thermal fluctuation if $x_{0}\gtrsim K$. Although the
mechanism of ParA localization could depend on the species, the estimate
suggests that the force plays a significant role in the partitioning
of plasmid/chromosome, independent of the specific molecular mechanisms.
Further, we note that the above-mentioned {}``spring toy model''
under a gradient is applicable to the partitioning problem and may
help explain the bifurcation of plasmid/chromosome in cell division
with the change in ParA localization. 

 Of course, the biological ParA distribution is more complicated.
Although the application mentioned here assumes ParA steady distribution, the distribution sometimes shows pole-to-pole oscillation \cite{F-plasmid-positioning-2} and regular patterns to form regular spacing of a few plasmid foci \cite{F-plasmid-positioning-1,F-plasmid-positioning-1.5}.
Still, our theory is applicable to the time-varying distribution, because the force derived here leads to directional motion toward the ParA foci.
 Indeed, the directional motion toward ParA focus has been suggested in \cite{F-plasmid-positioning-2}.
 Furthermore, the present gradient force is applicable to reaction-diffusion system so that the plasmid is located according to the chemical concentration pattern. 
 
\paragraph*{Discussion }

In this study, we have demonstrated that when a chemical concentration
gradient exists, a macroscopic element that acts as a scaffold for
the absorption of some chemical is generally subject to a thermodynamic
force, and it moves in the direction of increasing chemical potential.
The force is generated in accordance with the second law of thermodynamics,
as represented by the decrease in the grand potential in thermodynamics
under a given gradient; thus, this force acts independent of the specific
molecular mechanism. 
The gradient force is expected to provide a novel perspective on mechano-chemical
coupling. We have derived a general formula for the magnitude of the
force and presented a few conditions necessary for this formula to
be valid. The conditions are shown to be satisfied for the intracellular
motion of macromolecules or organelles under a suitable chemical gradient,
and the force is clearly shown to be larger than the thermal fluctuation
force. To be specific, we have discussed the application of this formula
to bacterial partitioning systems in cell division.
Considering the generality of our formulation, the chemical gradient force is
expected to be a universal guiding principle in intracellular organization.

The authors would like to thank H.Takagi, A.Awazu, S.Ishihara, and
T.Yomo for discussions.

\newpage 
\paragraph*{}
\newpage

\appendix*{\bf Supplementary information \rm }
\paragraph*{}
\paragraph*{Proof of the variational principle for grand potential with the principle of maximum work --- }

It is well known that the grand potential is a thermodynamic potential
in terms of natural variables $T,\mu,$ and surface area; extensive
studies such as those on adsorption and capillary phenomenon involve
this potential. However, to the best of our knowledge, there is no
proof of the variational principle for the grand potential, particularly
when $\mu$ varies; nevertheless, this potential is closely connected
by the Legendre transformation to the other thermodynamic potentials
for which the existence of variational principles has been proven.
It seems to be because
\begin{itemize}
\item standard thermodynamic operation involves extensive variables but
not intensive ones in order to obtain work from a thermodynamic system.
\end{itemize}
This is not the case in this study because $\mu$ can be controlled
by changing the $d$-dimensional position of the thermodynamic system
($d$=1,2,3). Therefore, we will derive the variational principle
for the grand potential on the basis of the principle of maximum work.

Consider a thermodynamic system composed of a bead with a surface
as shown in (Fig.1
). 
The bead is placed at $\boldsymbol{r}=\boldsymbol{\xi}$ and moves
in a $d$-dimensional space $\boldsymbol{r}\in\boldsymbol{R}^{d}$
($d$=1,2,3). It makes contact with a heat-particle bath (reservoir)
with homogeneous temperature $T(\boldsymbol{r})=T$ and chemical potential
$\mu(\boldsymbol{r})$ (concentration $x(\boldsymbol{r})$) at $\boldsymbol{r}=\boldsymbol{\xi}$
for a chemical X. The chemical is adsorbed on the surface of the bead
B and forms a complex Y with it (Fig.1
):$\hspace{.05in}{\rm X(\boldsymbol{\xi})+B\leftrightarrows Y}$.
The concentration on the bead is denoted by $y$.

Assume that the bead is in equilibrium with the reservoir at $\boldsymbol{r}=\boldsymbol{\xi}$
and it is maintained in equilibrium during its motion. Therefore,
the temperature and chemical potential of the system are always balanced
by those of the reservoir, that is, $T_{system}=T,\mu_{system}=\mu(\boldsymbol{\xi})$.
Note that we assume the existence of local equilibrium, where several
thermodynamic variables can be defined in the $d$-dimensional space.

The precise definition of the grand potential is as follows: \[
{\displaystyle \Omega(T,\mu):=\min_{y}\left[F(T,y)-y\mu\right]}\]
 This is the precise representation for the Legendre transformation
from $F(T,y)$, the Helmholtz free energy in terms of natural variables
$(T,y)$, to $\Omega(T,\mu)$ in terms of natural variables $(T,\mu)$;
mathematically, this implies that $y$ must be determined as a function
of $T,\mu$, so that $F(T,y)-y\mu$ is minimized when $y$ is varied
with fixed $T,\mu$. The determined $y^{*}$ and $F(T,y^{*})-y^{*}\mu$
are identical to the Langmuir isotherm and the grand potential, respectively.

The principle of maximum work represents the second law of thermodynamics
in terms of the work obtained by an isothermal process. First, we
assume that the principle of maximum work is applicable, and we vary
$y$ while keeping $T$ constant: \[
W_{max}(T,y_{1}\rightarrow y_{2})=F(T,y_{1})-F(T,y_{2})\]

When $y$ is quasistatically changed from $y_{1}$ as a function of
$\mu_{1}$ to \em any \rm $y_{2}$ by moving the position of the bead
from $\boldsymbol{\xi}_{1}$ to $\boldsymbol{\xi}_{2}$ via the force
$\boldsymbol{f}$ exerted by the external world (This force was balanced
by the chemical gradient force or forces from the other potential
fields, such as mechanical or magnetic forces.), the work $W(T,\mu_{1}\rightarrow\mu_{2})$
in terms of natural variables $T,\mu$ obtained by changing $\mu$
from $\mu_{1}$ to $\mu_{2}$ and keeping $T$ fixed, which is the
work performed by the system on the external world and is given as
follows:

\[
W(T,\mu_{1}\rightarrow\mu_{2})=-\int_{\boldsymbol{\xi}_{1}}^{\boldsymbol{\xi}_{2}}\boldsymbol{f}\cdot d\boldsymbol{\xi}\]
 \[
=W_{max}(T,y_{1}\rightarrow y_{2})-\left(-\int_{y_{1}\mu_{1}}^{y_{2}\mu_{2}}d\left(y\mu\right)\right)\]

The first term represents the total work that the system can perform,
and the second term represents the work that it must perform on the
reservoir for the change $y_{1}\rightarrow y_{2}$, i.e., the work
that it cannot perform on the external world. We can interpret the
latter work as follows: the potential energy of the system is $-y(\boldsymbol{\xi})\mu(\boldsymbol{\xi})$
at $\boldsymbol{r}=\boldsymbol{\xi}$ when it is placed in a potential
field $\mu(\boldsymbol{r})$, similar to a situation in which a dipole
is placed in the magnetic field generated by a magnet, as discussed
below. Furthermore, the former (latter) work is related to the interactions
that result in heat (chemical X) exchange with the reservoir, respectively.
The interactions are represented by changing $y$ through $\boldsymbol{\xi}$
while $T$ remains fixed. The essential point is that only the latter
work yields information on the system as an open system.

We consider maximizing $W(T,\mu_{1}\rightarrow\mu_{2})$ by changing
$y_{2}$ determined arbitrarily so far and then determining it as
a function of $T,\mu_{2}$. The maximum work is expressed in terms
of natural variables $T,\mu$ as $W_{max}(T,\mu_{1}\rightarrow\mu_{2})$.
We obtain the maximum work in terms of natural variables $(T,\mu)$
as follows: \[
W_{max}(T,\mu_{1}\rightarrow\mu_{2})={\displaystyle \max_{y_{2}}{\textstyle W(T,\mu_{1}\rightarrow\mu_{2})}}\]
 \[
={\displaystyle \max_{y_{2}}{\textstyle \left[W_{max}(T,y_{1}\rightarrow y_{2})+y_{2}\mu_{2}-y_{1}\mu_{1}\right]}}\]
 \[
=F(T,\mu_{1})-y_{1}\mu_{1}-{\displaystyle \min_{y_{2}}{\textstyle \left[F(T,\mu_{2})-y_{2}\mu_{2}\right]}}\]
 \[
=\Omega(T,\mu_{1})-\Omega(T,\mu_{2})\]
 That is, the maximum work that the system assigned $(T,\mu(\boldsymbol{\xi}))$
by the reservoir $(T,\mu(\boldsymbol{r}))$ at $\boldsymbol{r}=\boldsymbol{\xi}$
can perform on the external world is obtained by calculating the decrease
in the grand potential $\Omega(T,\mu(\boldsymbol{\xi}))$.

Next, we consider the situation in which the force $\boldsymbol{f}$
exerted by the external world vanishes ($\boldsymbol{f}=\boldsymbol{0}$)
and the change $\mu\rightarrow\mu'$ occurs spontaneously. In this
case, $W(T,\mu\rightarrow\mu')=0$ because $\boldsymbol{f}=\boldsymbol{0}$.
\[
\therefore W_{max}(T,\mu\rightarrow\mu')\geq W(T,\mu\rightarrow\mu')=0\]
 \[
\Leftrightarrow\Omega(T,\mu')-\Omega(T,\mu)\leq0\]

Because the start state $(T,\mu)$ can be chosen arbitrarily, the
grand potential monotonically decreases till the end state $(T,\mu')$
under local equilibrium is reached. If $(T,\mu')$ is in equilibrium,
the grand potential reaches a minimum at this state.

For an infinitesimal displacement $\boldsymbol{\xi}\rightarrow\boldsymbol{\xi}+d\boldsymbol{\xi}$,
the following inequality is satisfied:

\[
d\Omega(T,\mu)=\boldsymbol{\nabla}\Omega(T,\boldsymbol{\xi})\cdot d\boldsymbol{\xi}<0\]

\paragraph*{}

Therefore, for a displacement per unit time $\dot{\boldsymbol{\xi}}$,
\[
\dot{\Omega}(T,\boldsymbol{\xi},\dot{\boldsymbol{\xi}})=\boldsymbol{\nabla}\Omega(T,\boldsymbol{\xi})\cdot\dot{\boldsymbol{\xi}}<0\]
 Because the change in the grand potential is completely dissipated
due to $W(T,\mu\rightarrow\mu+d\mu)=0$ if the kinetic energy of the
bead is negligible, the following inequality is satisfied: \[
-\dot{\Omega}(T,\boldsymbol{\xi},\dot{\boldsymbol{\xi}})=T\sigma(\dot{\boldsymbol{\xi}})>0\]
 $\sigma(\dot{\boldsymbol{\xi}})$ is the entropy production of the
system. Thus, we can show that the variational principle applies to
the grand potential of the system.

If the dissipation in the system is only caused by friction due to
the motion of the bead, the following relation is satisfied for the
power that the system does for friction $\dot{W}_{fr}(\dot{\boldsymbol{\xi}})$.
\[-\dot{\Omega}(T,\boldsymbol{\xi},\dot{\boldsymbol{\xi}})=T\sigma(\dot{\boldsymbol{\xi}})=\dot{W_{fr}}(\dot{\boldsymbol{\xi}})>0\]

\paragraph*{Comparison with Gibbs free energy of magnetic dipole moment --- }

For understanding the work that the system must perform on the reservoir
as discussed above, it may be convenient to compare the problem with
that of a magnetic dipole moment placed in a magnetic field.

It is well known that a force exerted on the dipole placed in a magnetic
field acts in the direction of the larger one when the strength of
the magnetic field is inhomogeneous. The Gibbs free energy of the
magnetic dipole moment in terms of natural variables $T,H$, for which
its demagnetizing field is negligible, is defined as follows: \[
{\displaystyle G(T,H):=\min_{M}[F(T,M)-MH]}\]
 $F(T,M),H,M$ are the Helmholtz free energy of the dipole with natural
variables $T,M$, magnetic field, and magnetization per unit volume,
respectively. With $T$ fixed, \[ dG=dF-d(MH)=-MdH=-M\boldsymbol{\nabla}H\cdot d\boldsymbol{\xi}\]
 Now we will show that $dG(T,H)<0$, similar to the above discussion
on that of the grand potential. When the dipole at $\boldsymbol{r}=\boldsymbol{\xi}$
is placed in the magnetic field $H(\boldsymbol{r})$, it has potential
energy $-M(\boldsymbol{\xi})H(\boldsymbol{\xi})$. This means that
under an infinitesimal displacement $\boldsymbol{\xi}\rightarrow\boldsymbol{\xi}+d\boldsymbol{\xi}$,
the system must perform work $-d\left(M(\boldsymbol{\xi})H(\boldsymbol{\xi})\right)$
on the magnetic field in addition to the work performed on the external
world. It is clear that this is equivalent to $-d\left(y(\boldsymbol{\xi})\mu(\boldsymbol{\xi})\right)$,
when $y,\mu$ are replaced with $M,H$ in the above discussion. Therefore,
we see that when $T$ is fixed, $dG(T,H)<0$. 
When there is dissipation due to friction, assumed to be proportional to the velocity with the proportionality constant $\gamma$, the equation of motion for the dipole is given as follows: 
\[\gamma\dot{\boldsymbol{\xi}}=M\boldsymbol{\nabla}H\]

\end{document}